\documentclass{iau} 
\usepackage{graphicx}
\usepackage{hyperref}
\usepackage{commath}

\title[] 
{Strong FeII emission in NLS1s: An unsolved mystery}

\author[S. Panda, K. Ma\l{}ek, M. {\'S}niegowska \& B. Czerny]   
{\href{panda@cft.edu.pl}{Swayamtrupta Panda}$^{1,2}$; Katarzyna Ma\l{}ek$^{3,4}$; Marzena {\'S}niegowska$^2$ \and Bo{\.z}ena Czerny$^1$}

\affiliation{$^1$ Center for Theoretical Physics - PAS, Al. Lotnik{\'o}w 32/46, Warsaw, Poland 02-668 \\
$^2$  Nicolaus Copernicus Astronomical Center, ul. Bartycka 18, Warsaw, Poland 00-716 \\
$^3$ National Center for Nuclear Research, ul. Ho{\.z}a 69, Warsaw, Poland 00-681 \\
$^4$ Aix Marseille Univ. CNRS, CNES, LAM Marseille 13388, France}
\pubyear{2018}
\volume{341}  
\jname{PanModel2018 Osaka}
\editors{M. Boquien, E. Lusso, C. Gruppioni, \& P. Tissera, eds.}
\begin{document}

\maketitle

\begin{abstract}
The work initially started as a test to retrace the \cite{sh14} Quasar Main Sequence diagram where they claimed that the parameter R$\mathrm{_{FeII}}$, which defines the Eigenvector 1 (EV1) is driven by the Eddington ratio alone (\cite{pan18a}). We subsequently construct a refined (error and redshift limited) sample from the original \cite{shen11} QSO catalog. Based on our hypothesis - the main driver of the Quasar Main Sequence is the  maximum of the accretion disk temperature (T$\mathrm{_{BBB}}$) defined by the Big Blue Bump on the Spectral Energy Distribution (\cite{pan17}, \cite{pan18b}). We select the four extreme sources that have R$\mathrm{_{FeII}} \geq $ 4.0 and use the SED modelling code \href{https://cigale.lam.fr/}{CIGALE} (\cite{noll09}; \cite{boq18}) to fit the multi-band photometric data for these sources. We also perform detailed spectral fitting including the Fe II pseudo-continuum (based on the fitting parameters used in our paper \cite{snieg18}) to estimate and compare the value of R$\mathrm{_{FeII}}$ for them. We show the metallicity dependent FeII strength in the context of this study.
\end{abstract}

\firstsection 

\section{Introduction}
\cite{sh14} presented a $\mathrm{FWHM(H\beta)} - \mathrm{R_{FeII}}$ diagram for $\sim$ 20,000 SDSS quasars using the \cite{shen11} QSO catalog. To clean this diagram we added criteria selecting only objects with  $<$20\% error in the full-width at half maximums (FWHMs) and the equivalent widths (EWs) for the FeII (integrated) and H$\beta$ lines). This gives us our base sample with 4989 quasars. The FeII strength (R$_{\mathrm{FeII}}$= EW$_{\mathrm{FeII}}$/EW$_{\mathrm{H\beta}}$) where the FeII emission is considered within 4434-4684 \AA$\;$ (\cite{bor92}). Selecting sub-samples with R$_{\mathrm{FeII}} \geq$ 4.0, gives us 5 sources (see \cite{pan18a}) of which 1 is not a Narrow-line Seyfert 1 AGN and hence excluded in this study. 
\par 
\section{CIGALE SED analysis}
\begin{figure}
    \centering
    \includegraphics[width=\textwidth]{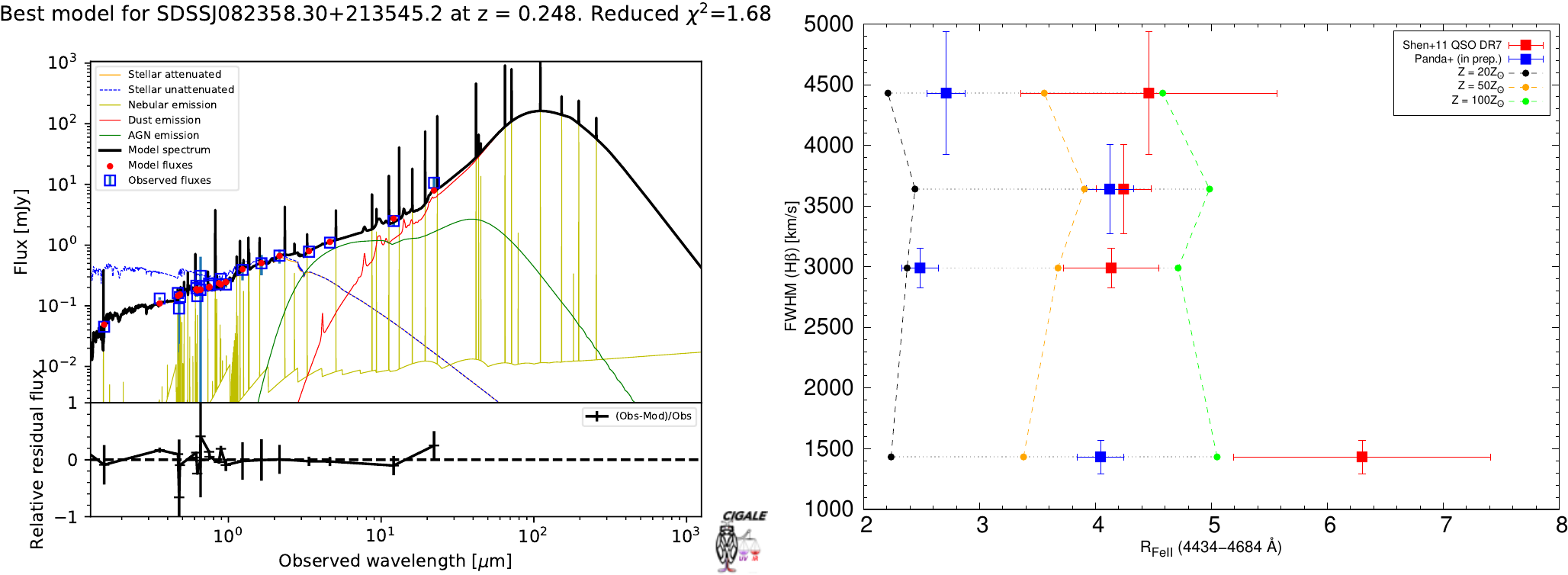}
    \caption{(a) CIGALE model fitting for SDSSJ082358.30+213545.20 - the orange line represents the stellar (attenuated) emission, the blue dashed line is for the stellar (unattenuated), the nebular, dust, AGN emission are shown in light green, red and dark green respectively. The overall model spectrum is shown in black with the observed (blue boxes) and model (red points) fluxes. (b) The plot shows a comparison between values estimated by an automatic template fitting procedure (red squares from \cite{shen11} QSO catalogue) versus our semi-automatic line-fitting algorithm (blue squares). A subset of the photoionisation grid simulation results is shown in the background for varying metallicities.}
    \label{fig2}
\end{figure}

We use CIGALE to fit multi-wavelength photometric data (5GHz to 1344 \AA). We utilise: (a) SFH with a delayed + exponential burst; (b) SSP: \cite{bc03}; (c) nebular emission; (d) \cite{cal00} dust attenuation; (e) \cite{dl2014} dust emission; and (f) \cite{fritz06} AGN model.  With the inclusion of optical-UV data, we were able to model the Big Blue Bump feature in the SED (spectral energy distribution), although the current setup underestimates the AGN contribution and compensates with a higher young stellar population (see Fig.\ref{fig2} left panel). CIGALE's SED analyses shows good agreement with the $\mathrm{T}_{\mathrm{BBB}}$ derived from the observations (see Table 1\footnote[2]{the table and the model fits for the other 3 sources can be found \href{https://drive.google.com/file/d/1At2NI8Jvl67dIDwV7m9AXFYSfKv7wdT0/view?usp=sharing}{here}}).



\section{Photoionisation Predictions On Metallicity}
We sought to predict the behaviour of these ``strong FeII'' emitting NLS1s (see Fig.\ref{fig2} right panel) using SEDs from observation. We use the photoionisation code CLOUDY (\cite{fer17}) to predict the line luminosities for FeII (integrated) and H$\beta$. In \cite{pan18b}, we have shown that the FeII emitting regions occupy the dark side of the broad line region (BLR) clouds which is also consistent with the time lags from recent reverberation mapping surveys. We found that these BLR clouds have high density ($\mathrm{n}_{\mathrm{H}} = 10^{12}\;\mathrm{cm}^{-3}$) and characteristic FeII emission can be modelled using high column density ($\mathrm{N}_{\mathrm{H}} = 10^{24}\;\mathrm{cm}^{-2}$) and a low value of (micro)turbulence ($\mathrm{v}_{\mathrm{turb}} = 10\;\mathrm{km/s}$). FeII strength (R$\mathrm{_{FeII}}$) estimates from photoionisation are further confirmed incorporating a high value of metallicity. This raises more questions toward the formation of the FeII in these AGNs. We performed a grid simulation that gave us a power-law dependence of the FeII strength on the metallicity:
\begin{equation}
    \mathrm{\langle log\left(R_{FeII} \right )\rangle}= -(0.325\pm 0.0261) + (0.519\pm 0.008)\mathrm{\langle log\left ( \frac{Z}{Z_{\odot}}\right )\rangle}\footnote[1]{average values based on the 4 sources}
\end{equation}
The predicted average metallicities are in the range of 20-55 $\mathrm{Z}_{\odot}$. Such high values of metallicities have been obtained for some high-redshift quasars with similar emission line features.

\section{The Big Question?}
CIGALE SED analyses for a small sample of extreme sources have shown good agreement for the values estimated for the $\mathrm{T}_{\mathrm{BBB}}$ with respect to observed numbers, although an AGN model extending to optical-UV is needed. With a broadband photometric model fitting we would like to investigate the contribution of the dust species around the central nuclei. A connection between the stellar environments and the AGN if established, can help solve two important questions (A) \textit{How are such ``underdeveloped'' galaxies harbouring relatively small but active black holes have such high FeII production?}, and (B) \textit{How the FeII arrived in the BLR?}''


\end{document}